\documentclass[showpacs,twocolumn,aps,prl]{revtex4}

\usepackage{amsmath,amssymb}
\usepackage{graphicx}
\usepackage{pst-plot}

\begin{document}

\title{Failure of Bell's Theorem and the Local Causality of the Entangled Photons}

\author{Joy Christian}

\email{joy.christian@wolfson.ox.ac.uk}

\affiliation{Department of Physics, University of Oxford, Parks Road, Oxford OX1 3PU, United Kingdom}

\begin{abstract}
A counterexample to Bell's theorem is presented which uses a pair of photons instead of spin-1/2 particles
used in our previous counterexamples. A locally causal protocol is provided for Alice and Bob, which
allows them to simulate observing photon polarizations at various angles, and record their results as
${A=\pm\,1\in S^3}$ and ${B=\pm\,1\in S^3}$, respectively. When these results are compared, the correlations
are seen to be exactly those predicted by quantum mechanics; namely ${\cos2(\alpha - \beta)}$, where ${\alpha}$
and ${\beta}$ are the angles of polarizers. The key ingredient in our counterexample is the topology of
3-sphere, which remains closed under multiplication, thus preserving the locality condition of Bell.
\end{abstract}


\maketitle

One of the first steps we often take towards measuring a physical quantity is to set up a Cartesian
coordinate system ${\{x,y,z\}}$ in the Euclidean space ${{\mathbb E}_3}$. This amounts to
modeling the Euclidean space as a 3-fold product of the real line, ${{\rm I\!R}^3}$. This
procedure has become so familiar to us that in practice we often identify ${{\mathbb E}_3}$
with its Cartesian model, and simply think of ${{\rm I\!R}^3}$ as {\it the} Euclidean
space. As we shall see, however, this seemingly innocuous act of convenience comes with a
very heavy price: It is largely responsible for the ${\,}$illusions${\,}$ of ``quantum non-locality.''
Once a coordinate-free geometric model of the Euclidean space is used, the correlations observed in
the EPR-type experiments involving photon pairs \cite{Peres-1993}, namely
\begin{align}
A(\alpha) = \pm\,1,\;B(\beta) = \pm\,1,\, \notag \\
{\cal E}(\alpha) = 0,\;\;\; {\cal E}(\beta) = 0,\;\;\;\, \notag \\
{\cal E}(\alpha,\,\beta) \,=\, \cos2(\alpha - \beta),\;\,\label{q}
\end{align}
are easily understood, in a {\it strictly}${\,}$ local-realistic terms.

Euclid himself of course did not think of ${{\mathbb E}_3}$ in terms of triples of real numbers.
He defined its representation axiomatically, entirely in terms of primitive geometric objects
such as points and lines, together with a list of their properties,
from which his theorems of geometry follow. Today we know, however, that it is quite tricky
to give a suitable definition of Euclidean space in the spirit of Euclid, and hence in physics
we instinctively identify ${{\mathbb E}_3}$ with ${{\rm I\!R}^3}$ whenever possible. But there is no
natural, geometrically-determined way to identify the two spaces without introducing an {\it unphysical}${\,}$
notion of arbitrarily distinguished coordinate system. This difficulty is clearly relevant in the
study of Bell's theorem \cite{Bell-1964}, for time and again we have learned that surreptitious introduction
of unphysical ideas in physics could lead to distorted views of the physical reality. A coordinate-free
representation of the Euclidean space is undoubtedly preferable, if what is at stake is the very
nature of the physical reality.

Fortunately, precisely such a representation of ${{\mathbb E}_3}$, with a rich algebraic structure, was provided
by Grassmann\break in 1844 \cite{Clifford}. As in Euclid's geometry, the basic elements of this powerful structure
are not coordinate systems, but points, lines, planes, and volumes, {\it all treated on equal footing}. Today one
begins this framework by postulating a unit volume element (or a trivector) in ${{\mathbb E}_3}$, defined by
\begin{equation}
I:={{\bf e}_x}\wedge\,{{\bf e}_y}\wedge\,{{\bf e}_z}\,,\label{2}
\end{equation}
with ${\{{\bf e}_x,\,{\bf e}_y,\,{\bf e}_z\}}$ being a set of orthonormal vectors in ${{\rm I\!R}^3}$, and ``${\wedge}$''
the ``outer'' product of Grassmann \cite{Clifford}. Each vector ${{\bf e}_j}$ is then a solution of
the equation ${I\wedge{\bf e}_j\,=\,0}$, and every pair of them respects the fundamental product
\begin{equation}
{\bf e}_j\,{\bf e}_k\,=\,{\bf e}_j\cdot{\bf e}_k+\,{\bf e}_j\wedge\,{\bf e}_k\,,\label{b-algebra}
\end{equation}
where ${{\bf e}_j\wedge\,{\bf e}_k}$ are unit bivectors, with counterclockwise sense for the cyclicly
permuted indices (${j,\,k=x,\,y,\;{\rm or}\;z}$). The resulting structure is a space spanned by the basis
\begin{equation}
\left\{1,\,{\bf e}_x,\,{\bf e}_y,\,{\bf e}_z,\,{\bf e}_x\wedge{\bf e}_y,\,
{\bf e}_y\wedge{\bf e}_z,\,{\bf e}_z\wedge{\bf e}_x,\,
{\bf e}_x\wedge{\bf e}_y\wedge{\bf e}_z\right\}\!,
\end{equation}
and encodes a graded linear algebra of dimensions eight. This algebra intrinsically characterizes the space ${{\mathbb E}_3}$.

Our interest, however, lies in a certain subalgebra of this algebra, the so-called even subalgebra
of dimensions four, defined by the bivector (or spinor) basis \cite{Eberlein}:
\begin{equation}
\left\{1,\,{\bf e}_x\wedge{\bf e}_y,\,
{\bf e}_y\wedge{\bf e}_z,\,{\bf e}_z\wedge{\bf e}_x\right\}\!.\label{5}
\end{equation}
Crucially for our purposes, this subalgebra happens to remain closed under multiplication. Consequently, it can\break
be used by
itself to model the Euclidean space ${{\mathbb E}_3}$. In fact, physically it provides the most consistent coordinate-free
representation of the Euclidean 3-space \cite{Eberlein}. The vectors and trivectors are then no longer intrinsic to the
basic subalgebra, but belong to a dual space. Only the scalars and bivectors---{\it treated on equal footing}---are taken to be
the intrinsic parts of the algebra. This can be seen more clearly if we use the condition ${I\wedge{\bf e}_j=0\,}$ to rewrite
the basis bivectors in equation (\ref{5}) as ${I\cdot{\bf e}_z}$, ${I\cdot{\bf e}_x}$, and ${I\cdot{\bf e}_y}$.
Their geometric product, analogous to equation (\ref{b-algebra}), then leads to the defining
equation of this subalgebra:
\begin{equation}
(I\cdot{\bf e}_j)\,(I\cdot{\bf e}_k)\,=\,-\;\delta_{jk}\,-\,\epsilon_{jkl}\;(I\cdot{\bf e}_l).\label{6}
\end{equation}
Evidently, despite the occurrences of trivectors and basis vectors, only the basis scalar and bivectors are involved in this
definition. Consequently, in what follows only scalars and bivectors (and their combinations) will have direct
physical significance---vectors and trivectors will merely facilitate computational ease, or ``hidden variables.''

Given the bivector basis defined by equation (\ref{5}), any generic bivector ${I\cdot{\bf a}}$ can be expanded in this basis as
\begin{equation}
I\cdot{\bf a}\,=\,
\{\,a_x\;{{\bf e}_y}\,\wedge\,{{\bf e}_z}
\,+\,a_y\;{{\bf e}_z}\,\wedge\,{{\bf e}_x}
\,+\,a_z\;{{\bf e}_x}\,\wedge\,{{\bf e}_y}\}.
\label{mu}
\end{equation}
It is worth stressing here that, although there is clearly isomorphism between the Euclidean vector space and the bivector
space, a bivector is an {\it abstract entity of its own}, with properties quite distinct from those of a vector \cite{Clifford}.
Given two such unit bivectors, say ${I\cdot{\bf a}}$ and ${I\cdot{\bf b}}$, the bivector subalgebra (\ref{6})
leads to the well known identity
\begin{equation}
(I\cdot{\bf a})(I\cdot{\bf b})\,=\,-\,{\bf a}\cdot{\bf b}\,-\,I\cdot({\bf a}\times{\bf b}),\label{i}
\end{equation}
provided we use the duality relation ${{\bf a} \wedge {\bf b}\,=\,I\cdot({\bf a}\times{\bf b})}$.

As we have discussed elsewhere \cite{topology}, the above identity provides a natural representation for the points of a unit
3-sphere. The bivectors ${I\cdot{\bf a}}$ and ${I\cdot{\bf b}}$, appearing on its L.H.S., represent the equatorial points of a
unit 3-sphere, and the {\it real} quaternion appearing on its R.H.S represents a non-equatorial point of the same sphere. An equator
of a 3-sphere, however, which of course is a 2-sphere, {\it does not} remain closed under multiplication. Therefore it is not
surprising that a product of any two points of a 2-sphere is not confined to the 2-sphere, but belongs to a 3-sphere. The
3-sphere itself, however, does remain closed under multiplication, thereby correctly encoding the topology underlying the
subalgebra (\ref{6}). Indeed, the R.H.S. of the above identity is not a pure bivector, but a sum of a scalar and a bivector.
And as we have discussed elsewhere \cite{topology}\cite{Can}, it is this {\it local-realistic}${\,}$ interplay between
the points of a 3-sphere and its equatorial 2-sphere---not the quantum entanglement---that is truly responsible for the EPR-type
correlations manifested in nature.

In particular, the fact that both the 3-sphere and its algebraic representation given by (\ref{i}) remain closed\break under
multiplication has profound implications for Bell's theorem. In any Bell-type experiment the measurement results of Alice
and Bob are usually taken to be numbers lying in some subset of the real line \cite{Bell-1964}. We have argued elsewhere
\cite{topology}\cite{Can}, however, that it is both physically and mathematically incorrect to regard measurement results
as points of the real line. In fact, unless they are taken to be the equatorial points of a 3-sphere, Bell's picture of the
physical reality would not be complete in the sense required by EPR. That is to say, unless the measurement results
are represented by maps of the form
\begin{equation}
A({\bf n},\,\lambda): {\rm I\!R}^3\!\times\Lambda\longrightarrow S^3, \label{map}
\end{equation}
the standard Bell-type prescription for a complete theory would be {\it necessarily incomplete}, and then there would
be no question of ``non-locality'' to begin with, because then Bell's argument simply would not get off the ground.
Indeed, {\it without completeness, there is no theorem}.

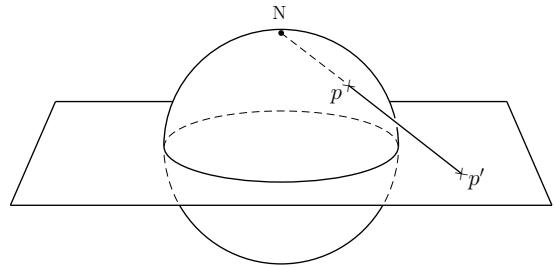
\begin{figure}
\hrule
\scalebox{0.6}{
\begin{pspicture}(-5.2,-3.7)(3.5,3.2)

\psarc[linewidth=0.2mm,linestyle=dashed](-1.0,-0.45){2.6}{184}{208}

\psarc[linewidth=0.2mm,linestyle=dashed](-1.0,-0.45){2.6}{332}{356}

\psarc[linewidth=0.3mm](-1.0,-0.45){2.6}{0}{9.6}

\psarc[linewidth=0.3mm](-1.0,-0.45){2.6}{14}{180}

\psarc[linewidth=0.3mm](-1.0,-0.45){2.6}{210}{330}

\psellipticarc[linewidth=0.17mm,linestyle=dashed](-1.0,-0.45)(2.6,0.8){0}{180}

\psellipticarc[linewidth=0.3mm](-1.0,-0.45)(2.6105,0.8){180}{360}

\psline[linewidth=0.3mm](-6.0,0.55)(-3.4,0.55)

\psline[linewidth=0.3mm](1.393,0.55)(4.0,0.55)

\psline[linewidth=0.3mm](-7.0,-1.75)(5.0,-1.75)

\psline[linewidth=0.3mm](-7.0,-1.75)(-6.0,0.55)

\psline[linewidth=0.3mm](4.0,0.55)(5.0,-1.75)

\psdot*(-1.0,2.07)

\rput{35}(3.0,-1.05){\large ${\times}$}

\rput{35}(0.5,0.904){\large ${\times}$}

\put(0.1,0.6){{\Large ${p}$}}

\put(-1.19,2.37){{\large ${\rm N}$}}

\put(3.15,-1.3){{\Large ${p'}$}}

\psline[linewidth=0.2mm,linestyle=dashed](-1.0,2.07)(0.39,1.0)

\psline[linewidth=0.3mm](0.5,0.904)(3.0,-1.05)

\end{pspicture}}
\hrule
\caption{Stereographic projection of ${S^2}$ onto the plane ${{\rm I\!R}^2}$. Both ${S^2}$ and ${{\rm I\!R}^2}$
contain infinite number of points. Each point ${p}$ of ${S^2}$ can be mapped to a point ${p'}$ of ${{\rm I\!R}^2}$, except
the north pole, which has no meaningful image under this projection.\break}
\label{fig}
\hrule
\end{figure}
The conceptual reasons for this are fairly elementary. Recall that the functions ${A({\bf n},\,\lambda)}$ postulated
by Bell are not only supposed to represent the measurement results, but also the end-points of a dynamical process within
a yet-to-be-discovered complete theory of physics. Indeed, as Bell himself puts it: ``In a complete physical theory of
the type envisaged by Einstein, the hidden variables would have dynamical significance and laws of motion; our ${\lambda}$
can then be thought of as initial values of these variables at some suitable instant'' \cite{Bell-1964}. Accordingly,
we may think of the functions ${A({\bf n},\,\lambda)}$ as solutions of some differential equation,
with ${\lambda}$ and ${\bf n}$ as the ``initial'' and ``final'' conditions, respectively. However one may think of them,
${A({\bf n},\,\lambda)}$ are {\it functions}, and as such they have a domain set, ${{\rm I\!R}^3\!\times\Lambda}$, from
which they take-in values, and a codomain set, ${\Sigma}$, to which they end up belonging:
\begin{equation}
A({\bf n},\,\lambda):
\left(\begin{array}{c}{\bf n}_1\\{\bf n}_2\\.\\{\bf n}_j\\.\\.\end{array}\right)\times
\left(\begin{array}{c}{\lambda}_1\\{\lambda}_2\\.\\.\\{\lambda}_k\\.\end{array}\right)
\longrightarrow\left(\begin{array}{c}A({\bf n}_1,\,\lambda_2)=+\,1\\A({\bf n}_2,\,\lambda_1)=-\,1\\.\\A({\bf n}_j,\,\lambda_k)
=+\,1\\.\\.\end{array}\right)\!.\label{like}
\end{equation}
The question then is: What is the codomain of ${A({\bf n},\,\lambda)}$? Usually it is assumed to be some subset
of ${{\rm I\!R}}$. Although na\"ive, this choice is good enough if one is concerned only with a finite number of
measurement results. But what if Alice decides to measure along all infinitely many of the possible directions?
After all, simultaneous existence of {\it all} elements of reality is central to the EPR argument. And a
theory that cannot accommodate every possible measurement result can hardly be considered complete.
It is however mathematically impossible to account for all possible measurement results by means of
${A({\bf n},\,\lambda)}$, unless its codomain is a 2-sphere. Here is the reason why:

In the standard EPR-Bell scenario there are infinitely many possible spin components that could be measured by Alice---one
corresponding to each direction ${{\bf n}\in{\rm I\!R}^3}$. Thus there is a one-to-one correspondence between the set of
measurement results obtainable by Alice and the points of a unit 2-sphere defined by ${||{\bf n}||=1}$.
As depicted in Fig.${\,}$1${\,}$ however, a 2-sphere is not homeomorphic to ${{\rm I\!R}^2}$ (or to ${\rm I\!R}$ for that
matter, for both ${\rm I\!R}$ and ${{\rm I\!R}^2}$ have the same cardinality). Thus it is topologically
impossible to account for every possible measurement result by means of a function like
(\ref{like}), unless its codomain
is a 2-sphere. Indeed, it is evident from Fig.${\,}$1 that if the codomain of ${A({\bf n},\,\lambda)}$ differs from 2-sphere
by even a single point, then the set of possible values it can take would fail to be in one-to-one correspondence with the set of
results Alice could obtain in principle. And then there would be at least one measurement result that would not have a counterpart
in a complete theory, rendering the function ${A({\bf n},\,\lambda)}$ worthless for the purposes of Bell (whose aim was to prove
that no {\it complete} theory can be locally causal).

Actually, even the 2-sphere does not quite capture the complete picture of local reality, because it fails to remain closed
under multiplication \cite{topology}. For any two-level system completeness and locality necessitate that the 2-sphere
be recognized as an equator of a 3-sphere \cite{topology}\cite{Can}. That is to say, for any two-level system
it is both necessary and sufficient
to use maps of the form (\ref{map}) for a complete local theory. Then not only the measurement results, but also their
products would remain within the 3-sphere---i.e., remain within the same topological space---respecting the locality
(or factorizability) condition of Bell. Conversely, given any arbitrary point ${P}$ of a 3-sphere, it can always be
factorized into any number of points: ${P=ABCD...}$ Needless to say, this is a highly nontrivial and powerful property
of the 3-sphere. To appreciate its non-triviality, consider a product of {\it infinitely many} points of a 3-sphere. Such
a product will simply be another point of the same sphere \cite{topology}. By contrast, this will not be true in the case
of a 2-sphere, even for just two points. As we shall see, it is this non-trivial property of the 3-sphere that sustains
the local causality of the entangled photons.

So far we have considered the bivector subalgebra with arbitrarily fixed basis as in definition (\ref{5}). The convention
usually is to assume a right-handed set of basis bivectors, and so far we have followed this convention. The algebra itself,
however, does not fix the handedness of the basis. We could have equally well started out with a left-handed set of bivectors,
by letting ${-\,I}$ instead of ${+\,I}$ fix the basis. Equation (\ref{6}) would have then had the alternate form:
\begin{equation}
(-I\cdot{\bf e}_j)\,(-I\cdot{\bf e}_k)\,=\,-\;\delta_{jk}\,-\,\epsilon_{jkl}\;(-I\cdot{\bf e}_l).
\end{equation}
Comparing this equation with equation (\ref{6}) we see that there
remains a sign ambiguity in the definition of our subalgebra ({\it cf}. Refs.${\,}$\cite{Clifford} and \cite{Eberlein}):
\begin{equation}
(I\cdot{\bf e}_j)\,(I\cdot{\bf e}_k)\,=\,-\;\delta_{jk}\,\pm\,\epsilon_{jkl}\;(I\cdot{\bf e}_l).\label{g}
\end{equation}
Consequently, following the time-honored mathematical practice of turning an ambiguity of sign into virtue, we define the
handedness of this entire algebra as a ``hidden variable'', just as in our previous counterexamples \cite{Christian}.
In other words, we specify the complete state of the photons we are about to study as ${\,{\boldsymbol\mu}=\pm\,I,\,}$
thereby defining the basis of our entire subalgebra by the equation
\begin{equation}
({\boldsymbol\mu}\cdot{\bf e}_j)\,({\boldsymbol\mu}\cdot{\bf e}_k)\,=
\,-\;\delta_{jk}\,-\,\epsilon_{jkl}\;({\boldsymbol\mu}\cdot{\bf e}_l).\label{14}
\end{equation}
The identity (\ref{i}) for the generic bivectors then becomes
\begin{equation}
(\,{\boldsymbol\mu}\cdot{\bf a})(\,{\boldsymbol\mu}\cdot{\bf b})\,
=\,-\,{\bf a}\cdot{\bf b}\,-\,{\boldsymbol\mu}\cdot({\bf a}\times{\bf b}),\label{bi-identity}
\end{equation}
along with the duality relation ${{\bf a}\wedge{\bf b}={\boldsymbol\mu}\cdot({\bf a}\times{\bf b})}$. In
other words, the duality between the wedge product and cross product within our subalgebra alternates
between the right-hand and left-hand rules \cite{Christian}. Clearly, then, our complete state ${{\boldsymbol\mu}=\pm\,I}$
represents a far deeper hidden structure than the simpleminded variables considered by Bell. It un-fixes the basis
of an entire algebra, and turns it into a shared randomness between Alice and Bob.

We are now well equipped---both conceptually and mathematically---to take up the question of EPR-Bohm
correlations exhibited by a pair of entangled photons. This question of course has been well scrutinized
in the literature.${\;}$We will restrict to the most basic aspects of the question, and follow its treatment
by Peres \cite{Peres-1993}.

In quantum mechanical description of the experiment involving photon polarizations one usually assumes that the
system has been prepared in the state
\begin{equation}
|\,\Psi_+\rangle\,=\,\frac{1}{\sqrt{2}\;}\Bigl\{|\,H\,\rangle_1\otimes|\,H\,\rangle_2\,
+\,|\,V\,\rangle_1\otimes|\,V\,\rangle_2\Bigr\}\,,\label{single}
\end{equation}
where ${|\,H\,\rangle}$ and ${|\,V\,\rangle}$ denote the horizontal and vertical polarization states of the photons
along the directions ${{\bf e}_x}$ and ${{\bf e}_y}$, respectively; and the subscripts 1 and 2 refer to the photons 1 and 2,
respectively. The photons are thus assumed to be propagating in the ${{\bf e}_z}$ direction. We could equally well
consider the singlet state ${|\,\Psi_-\rangle}$, but that would not add anything significant to our concerns here.
Both polarization states, ${|\,\Psi_+\rangle}$ and ${|\,\Psi_-\rangle}$, are invariant under rotations about the ${{\bf e}_z}$
axis, but the state ${|\,\Psi_+\rangle}$ is even under reflections, whereas the state ${|\,\Psi_-\rangle}$ is odd.

In a typical experimental run Alice and Bob measure polarizations of the photons along two different directions in the plane
perpendicular to the ${{\bf e}_z}$ axis. Alice measures polarizations along the direction ${\bf a}$, which makes an angle
${\alpha}$ with the ${{\bf e}_x}$ axis, whereas Bob measures polarizations along the direction ${\bf b}$, which makes an
angle ${\beta}$ with the ${{\bf e}_x}$ axis. Individually the binary results observed by Alice and Bob, namely
${A(\alpha) = \pm\,1}$ and ${B(\beta) = \pm\,1}$, are found to be random, with equal numbers of ${+\,1}$ and ${-\,1}$.
When these results are compared, however,
they are found to be correlated, in agreement with the quantum mechanical predictions we have summarized in equation (\ref{q}).

Our goal is to reproduce these quantum mechanical predictions {\it exactly}, within the local-realistic framework discussed
above (see also Refs.${\,}$\cite{topology} to \cite{Christian}).
To this end, we have let the complete state of the photons be
given by ${{\boldsymbol\mu}=\pm\,I}$, where ${I}$ is the basic trivector defined in equation (\ref{2}). The photon
polarizations observed by Alice and Bob along their
respective axes ${\bf a}$ and ${\bf b}$, for a given ``basis'' ${\boldsymbol\mu}$,
are then represented by the following two local beables:
\begin{align}
A(\alpha,\,{\boldsymbol\mu})&=+\,{\boldsymbol\mu}\cdot{\widetilde{\bf a}}\,,\;\;
{\widetilde{\bf a}}:= {\bf e}_x\,\sin2\alpha\,+\,{\bf e}_y\,\cos2\alpha\,, \\
B(\beta,\,{\boldsymbol\mu})&=-\,{\boldsymbol\mu}\cdot{\widetilde{\bf b}}\,,\;\;
{\widetilde{\bf b}}:= {\bf e}_x\,\sin2\beta\,+\,{\bf e}_y\,\cos2\beta\,.
\end{align}
To begin with, note that ${A(\alpha,\,{\boldsymbol\mu})}$ and ${B(\beta,\,{\boldsymbol\mu})}$ are strictly
{\it local} variables. Alice's measurement result depends only on her freely chosen angle ${\alpha}$ and the
initial state ${{\boldsymbol\mu}}$, and, likewise, Bob's measurement result depends only on his freely chosen angle
${\beta}$ and the initial state ${{\boldsymbol\mu}\,}$. Moreover, despite appearances, ${\,A(\alpha,\,{\boldsymbol\mu})}$
and ${B(\beta,\,{\boldsymbol\mu})\,}$ are simply binary outcomes, ${+1}$ or ${-1}$, albeit occurring within the
compact topology of a 3-sphere rather than the real line:
\begin{align}
A(\alpha,\,{\boldsymbol\mu})& \cong \pm\,1\in S^2\in S^3\;{\rm about}\;{\widetilde{\bf a}}\in{\rm I\!R}^3, \\
B(\beta,\,{\boldsymbol\mu})& \cong \pm\,1\in S^2\in S^3\;{\rm about}\;{\widetilde{\bf b}}\in{\rm I\!R}^3.
\end{align}
In fact, since the space of all ${{\boldsymbol\mu}\cdot{\widetilde{\bf a}}}$ is isomorphic to the equatorial 2-sphere
contained within a 3-sphere \cite{Zulli}, each measurement result ${A({\alpha},\,{\boldsymbol\mu})}$ of Alice---corresponding
to a preexisting element of reality of the photon---is uniquely identified with a definite point of this 2-sphere. In other words,
every preexisting polarization of a given photon gives rise to a unique measurement outcome ${A({\alpha},\,{\boldsymbol\mu})}$,
which in turn is unambiguously represented by a definite point of the equatorial 2-sphere \cite{topology}. Note also that,
since we are working entirely within the bivector model of the Euclidean space defined by (\ref{5}) and (\ref{14}),
parallel crystals of Alice and Bob will yield identical outcomes despite the apparent sign difference between
${A({\alpha},\,{\boldsymbol\mu})}$ and ${B({\beta},\,{\boldsymbol\mu})}$.

If we now assume that initially there was 50/50 chance between right-handed and left-handed orientations of the
Euclidean space---i.e., equal chance between the complete states ${{\boldsymbol\mu}=+\,I}$ and ${{\boldsymbol\mu}=-\,I}$,
then the expectation values of the above two points in ${S^2}$ trivially work out to be \cite{Christian}
\begin{equation}
{\cal E}(\theta)\,=
\lim_{n\rightarrow\,\infty}\left\{\frac{1}{n}\sum_{i\,=\,1}^{n}\, A({\theta},\,{\boldsymbol\mu}^i)\right\}\,=\;0\,,
\end{equation}
where ${\theta=\alpha}$ or ${\beta}$. On the other hand, the product of these two points
is a non-equatorial point lying within the 3-sphere, and hence it too would have a definite value:
\begin{equation}
A({\alpha},\,{\boldsymbol\mu})\,B({\beta},\,{\boldsymbol\mu}) \,\cong \pm\,1 \in S^3.
\end{equation}
In fact, our 3-sphere is entirely made of all such product points, each of definite value ${+\,1}$ or ${-\,1}$.
More precisely, the space of all products ${P=AB}$ is isomorphic to a unit 3-sphere,
${P_1^2+P_2^2+P_3^2+P_4^2=1}$ \cite{topology}\cite{Zulli}.${\;}$This
can be seen more clearly if we expand ${AB}$ using equation (\ref{bi-identity}):
\begin{align}
A({\alpha},\,{\boldsymbol\mu})\,B({\beta},\,{\boldsymbol\mu})&=
\{+\,{\boldsymbol\mu}\cdot{\widetilde{\bf a}}\,\}
\{-\,{\boldsymbol\mu}\cdot{\widetilde{\bf b}}\,\} \notag \\
&=+\,{\widetilde{\bf a}}\cdot{\widetilde{\bf b}}\,+
\,{\boldsymbol\mu}\cdot\left({\widetilde{\bf a}}\times{\widetilde{\bf b}}\right) \\
&=\cos2(\alpha-\beta) \,+\, ({\boldsymbol\mu}\cdot{\bf e}_z) \, \sin2(\alpha-\beta). \notag
\end{align}
Evidently, ${AB}$ describes a circle of points within ${S^3}$ \cite{Zulli}, each of definite value ${+\,1}$ or ${-\,1}$,
depending on ${\alpha}$, ${\beta}$, and ${\,{\boldsymbol\mu}\cdot{\bf e}_z\cong \pm\,1\in S^2\in S^3}$. With this parameterization
of the points of ${S^3}$,
the joint expectation value ${{\cal E}(\alpha,\,\beta)}$ of the measurement results of Alice and Bob is easily seen
to reproduce the quantum mechanical prediction:
\begin{align}
{\cal E}&({\alpha},\,{\beta})
=\lim_{n\rightarrow\,\infty}\left\{\frac{1}{n}\sum_{i\,=\,1}^{n}\, A({\alpha},\,{\boldsymbol\mu}^i)\;
B({\beta},\,{\boldsymbol\mu}^i)\right\} \notag \\
&=\,\cos2(\alpha-\beta)+\!\lim_{n\rightarrow\,\infty}\left\{\frac{1}{n}\sum_{i\,=\,1}^{n}\,
 ({\boldsymbol\mu}\cdot{\bf e}_z)^i \, \sin2(\alpha-\beta)\right\} \notag \\
&=\,\cos2(\alpha-\beta).
\end{align}
It is important to note here that the summation in both the first and the second line above is over the bivectors such as
${{\boldsymbol\mu}\cdot{\widetilde{\bf a}}}$, since we are working entirely within the bivector basis defined by equations
(\ref{5}) and (\ref{14}). Finally, as discussed in Ref.${\,}$\cite{Peres-1993}, the violations of CHSH inequality
follow trivially from the above expectation value: 
\begin{equation}
\left|\,{\cal E}({\alpha},\,{\beta})\,+\,{\cal E}({\alpha},\,{\beta}\,')\,+\,
{\cal E}({\alpha}\,',\,{\beta})\,-\,{\cal E}({\alpha}\,',\,{\beta}\,')\,\right|\,\leqslant\,2\sqrt{2}\,.
\notag \label{My-CHSH}
\end{equation}

This example is but one in a series of examples \cite{topology}\cite{Can}\cite{Christian} which show
that the illusion of ``quantum non-locality'' stems from a topologically incomplete accounting of
the elements of physical reality. The illusion evaporates when every possible measurement result is
correctly accounted for, in a topologically coherent manner. In our example this is achieved by modeling
the physical 3-space, not as ${{\rm I\!R}^3}$, but ${S^3}$, in terms of abstract coordinate-free geometry of
points and planes introduced by Grassmann \cite{Clifford}.

I wish to thank Daniel ben-Avraham, David Coutts, Daniel Rohrlich, and Jeffrey Uhlmann for discussions.

\end{document}